\documentclass[epj,nopacs]{svjour}
\usepackage{graphicx}
\usepackage{hyperref}
\usepackage{amsmath,amssymb}
\usepackage{cite}
\usepackage{soul}

\usepackage{color}
\definecolor{magenta}{rgb}{0.55, 0.0, 0.55}
\definecolor{orange}{rgb}{0.75, 0.25, 0.05}

\begin{document}
\title{Direct discovery of new light states at the FCCee}

\author{Simon Knapen\inst{1} \and Andrea Thamm\inst{2}}
%
%
\institute{CERN, Theoretical Physics Department, Geneva, Switzerland \and School of Physics, The University of Melbourne, Victoria 3010, Australia}
\date{Received: \today}
%
\abstract{
Light new states are ubiquitous in many models which address fundamental outstanding questions within the Standard Model (SM). The FCCee provides an excellent opportunity to probe these new particles with masses between $1$ and $100\,$GeV and their electroweak couplings. Here we discuss the theory motivations for axion-like particles and heavy neutral leptons and detail the potential of direct discovery at the FCCee. We highlight that our current understanding requires light new states to be embedded within a bigger theory framework and thus the complementarity of the precision frontier at the FCCee and the high energy frontier of the FCChh program.
%
} 
\maketitle
\section{Introduction}
\label{intro}

Interest in light new particles has gained traction in recent years and has led to exciting possibilities for experimental and theoretical progress. After several decades of a predominant focus on TeV-scale new physics, driven mostly by compelling solutions to the hierarchy problem, the lack of evidence of any new physics at the LHC has broadened our focus to a wider set of questions. While the search for physics at the TeV scale at the LHC remains as motivated as ever, theorists have started to (re)explore other ideas, many of which predict weakly coupled new particles with much lower masses.

Light new particles are particularly ubiquitous in theories which solve the hierarchy problem, such as supersymmetric and composite Higgs models, as well as models that address the strong CP problem, the origin of neutrino masses or dark matter, and leptogenesis. Here we discuss two well motivated examples: axion-like particles as heavy QCD axions and heavy neutral leptons as light right handed neutrinos. While the most minimal implementations of these models do not always predict light new states, they do occur in many reasonable and motivated extensions. The open parameter space is moreover still very large, and provides an excellent opportunity for the FCCee. 
In particular, the  FCCee will very precisely probe the possible existence of new particles with masses between $1$ and $100\,$GeV and their electroweak couplings. This mass range is difficult to access for the LHC and Tevatron due to trigger limitations and large backgrounds, while the particles are too heavy to be produced at Belle II.

A discovery of a light new state at the FCCee would be extremely exciting, as it would need to be embedded in a bigger framework that would require heavy new particles alongside it, which could hold answers to a wide possible range of open problems within the Standard Model (SM). While the FCCee can explore both weakly-coupled light new physics and sensitive electroweak observables with unparalleled accuracy, heavier states would be a clear target for the FCChh. This highlights the complementarity between the FCCee and FCChh programmes. Any discoveries or anomalies seen at the FCCee will very likely produce questions which require exploration of the high energy frontier to answer, which the FCChh can amply provide.

\section{Axion-like particles}

One of the major outstanding mysteries of the Standard Model is the fact that time-reversal or CP symmetry is violated in the Yukawa interactions of the quark sector, but nevertheless appears to be exceptionally well respected by the strong force. In particular, the QCD sector of the Standard Model is equipped with a hidden phase ($\theta$), whose non-zero value would imply additional CP violation.  Very precise measurements of the neutron electric dipole \cite{Abel:2020gbr} have however constrained $|\theta|\lesssim 10^{-13}$, which stands in sharp contrast with the $\mathcal{O}(1)$ CP-violating phase in the CKM matrix of the quark sector. One may certainly accept this so-called ``strong CP problem'' as an extreme coincidence in the parameters of the Standard Model, however such a tiny parameter, experimentally consistent with zero, does beg the question whether it is perhaps \emph{exactly} zero, and if so, why. 

\begin{figure}[t]\centering
\includegraphics[width =0.85\textwidth]{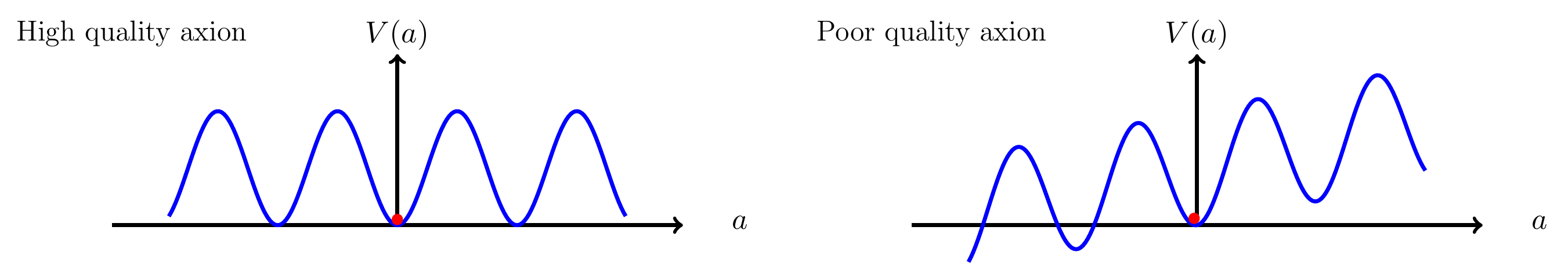}
\caption{(Left) Good quality axion, with the axion in a local minimum, CP conserving minimum at $\langle a\rangle=0$. (Right) Poor quality axion, with the axion in a local minimum, CP breaking minimum at $\langle a\rangle\neq0$. The red dot indicates the local minimum of the potential and thus $\theta=\langle a\rangle$\label{fig:axionquality}}
\end{figure}

In a set of brilliant papers in 1977 Roberto Peccei and Helen Quinn \cite{Peccei:1977hh,Peccei:1977ur} argued that $\theta$ could be the vacuum expectation of a \emph{dynamical} field, the \emph{axion} ($a$). They showed that the axion field has a potential
\begin{equation}\label{eq:axionpot}
V(a)=m_\pi^2 f_\pi^2\left[1-\cos\left(\frac{a}{f_a}\right)\right]
\end{equation}
with $m_\pi = 135\,$MeV denoting the pion mass and $f_\pi$ and $f_a$ the pion and axion decay constants, respectively. If this is the case, $a$ settles at $\theta=\langle a\rangle=0$ simply by minimizing its energy. This concept is shown schematically in the left-hand panel of Fig.~\ref{fig:axionquality}. The mass of $a$ can be found by expanding $V(a)$ in small $a$, $m_a=m_\pi f_\pi/f_a$.
As with most elegant solutions to fundamental puzzles, there is an important subtlety however: The solution of Peccei and Quinn only works if all other contributions to $V(a)$ are much smaller than those in \eqref{eq:axionpot}. If this is not the case, and the potential gets tilted even a tiny bit near the origin, the minimum will no longer occur at $\langle a\rangle=0$, reintroducing the strong CP problem (see right-hand panel in Fig.~\ref{eq:axionpot}). Quantum gravity in particular poses a problem, as it is expected to induce terms of the form $\sim a f_a^{\Delta-1} /M_{pl}^{\Delta-4}$, with $M_{pl} = 2.4\times 10^{18}\,$GeV the Planck mass and $\Delta$ a model dependent exponent. The solution appears obvious: We can simply restrict ourselves to parameter space where $f_a\ll M_{pl}$ to suppress these dangerous contributions. Quantitatively however, this implies $f_a\lesssim 10\,$TeV and $m_a\lesssim$ keV for the case of $\Delta=6$. Unfortunately, this is firmly excluded already by stellar cooling bounds \cite{Raffelt:1985nk,Raffelt:1987yu} and exotic Kaon transitions \cite{Lai:2002kf,Artamonov:2005ru,Abouzaid:2008xm,Ceccucci:2014oza}. This conundrum is known as the \emph{axion quality problem}. 

One possible solution is that the relation between $m_a$ and $f_a$ is in fact not as strict as predicted by \eqref{eq:axionpot}: There may be a dark sector which also contributes to $V(a)$, and thus acts to increase $m_a$, evading current experimental bounds~\cite{Rubakov:1997vp}. Models such as~\cite{Rubakov:1997vp} motive the search for axions with masses in the GeV to TeV range, which fortuitously is exactly the mass range that can be probed by future colliders. The fact that axions always couple to SM gluons should not come as a surprise given that this coupling defines the vanishing CP violating phase $\theta$ via ${\cal L}_\theta = \theta g_s^2/32\pi^2 G_{\mu\nu}^a\,\tilde G^{\mu\nu,a}$, where $g_s$ is the strong coupling constant. In grand unified models, the condition that the axion sector does not spoil the unification of the SM gauge couplings also predicts axion couplings to the electroweak gauge bosons. The extraordinary number of electroweak gauge bosons produced at the FCCee therefore makes it very sensitive to heavy axion models. An additional coupling to SM fermions is present in the class of models proposed by Dine, Fischler, Srednicki and Zhitnitsky (DFSZ) \cite{DINE1981199,Zhitnitsky:1980tq} but absent in models following the Kim-Shifman-Vainshtein-Zakharov \cite{PhysRevLett.43.103,SHIFMAN1980493} scheme.

Particles in this mass range and with very similar properties also occur in models which address the hierarchy problem rather than the strong CP problem. This fundamental problem refers to the fact that if there are any new, heavy particles in nature, then the electroweak scale naturally receives corrections roughly equal to their mass. Since the Standard Model is incomplete and is properly considered an effective theory, this raises the question as to why electroweak symmetry is being broken so many orders of magnitude below e.g.~the scale at which grand unification is expected to take place. This conclusion is avoided if a symmetry leads to an intricate cancellation or if the Higgs boson were actually a composite particle. Large classes of models have been built around these ideas and are referred to as \emph{supersymmetric} and \emph{composite Higgs} models. Particles similar to the axion can emerge in these models when a global symmetry is broken and further motivate our search for them. R-axions in supersymmetric models \cite{Bellazzini:2017neg} and light pseudoscalar particles in composite Higgs models \cite{Ferretti:2013kya} motivate couplings to both gauge bosons and fermions. 

Independently of their origin, these particles are referred to as {\it axion-like particles} (ALPs). Given this broad range of motivations for ALPs, it is very useful to consider an effective Lagrangian which parameterizes these models: 
\begin{equation}\label{Leff}
\begin{aligned}
   {\cal L}
   =& \frac12 \left( \partial_\mu a\right)\!\left( \partial^\mu a\right) - \frac{1}{2}m_{a}^2\,a^2 + \sum_f \frac{c_{ff}}{2} \frac{\partial^\mu a}{\Lambda}\, \bar f\gamma_\mu \gamma_5 f + c_{GG}\,g_s^2\,\frac{a}{\Lambda}\,G_{\mu\nu}^a\,\tilde G^{\mu\nu,a} + c_{WW}\,\frac{e}{\sin(\theta_w)}\,\frac{a}{\Lambda}\,W_{\mu\nu}\,\tilde W^{\mu\nu}
\\&
    + c_{\gamma\gamma}\,e^2\,\frac{a}{\Lambda}\,F_{\mu\nu}\,\tilde F^{\mu\nu}
    + c_{\gamma Z}\,\frac{4e^2}{\sin(2\theta_w)}\,\frac{a}{\Lambda}\,F_{\mu\nu}\,\tilde Z^{\mu\nu}
    + c_{ZZ}\,\frac{4e^2}{\sin^2(2\theta_w)}\,\frac{a}{\Lambda}\,Z_{\mu\nu}\,\tilde Z^{\mu\nu}
\end{aligned}
\end{equation}
with $g_s$, $e$ the strong and electromagnetic couplings respectively. $\theta_w$ is the Weinberg angle and $\Lambda$ is proportional to the axion decay constant $f_a$. $c_{ff}$, $c_{GG}$, $c_{WW}$, $c_{\gamma\gamma}$, $c_{\gamma Z}$ and $c_{ZZ}$ are model dependent parameters. 

\begin{figure}[t]\centering 
\includegraphics[width =0.2\textwidth]{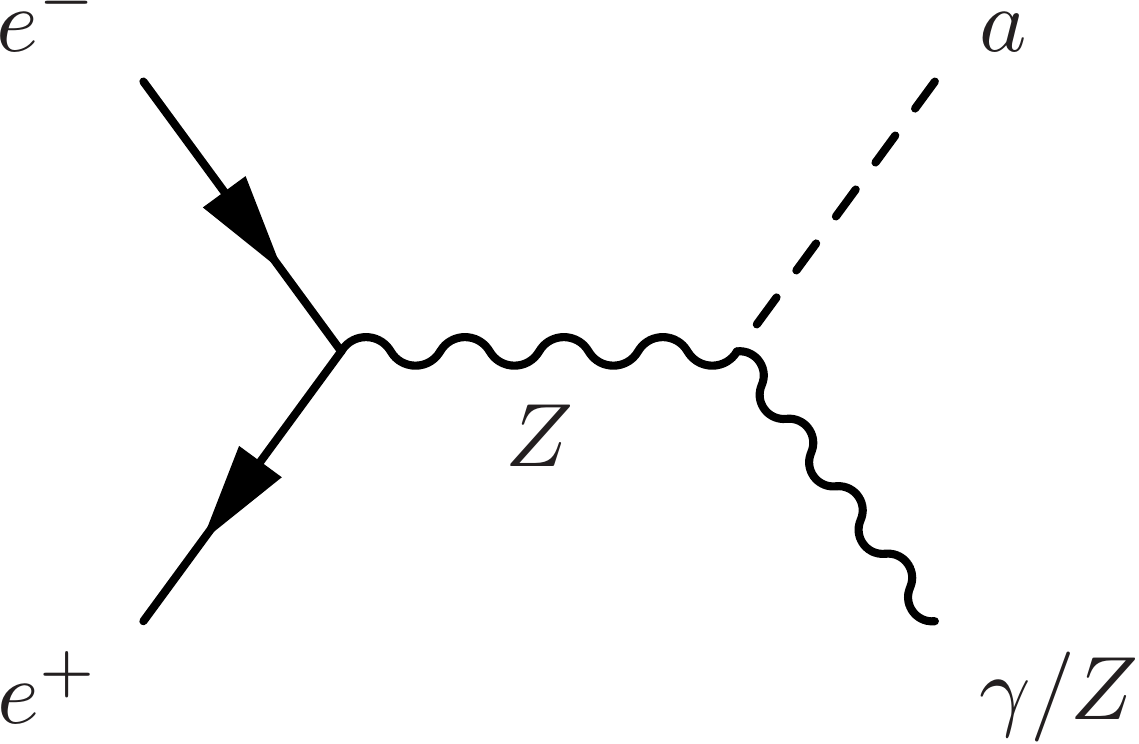} \hspace{2cm}
\includegraphics[width =0.2\textwidth]{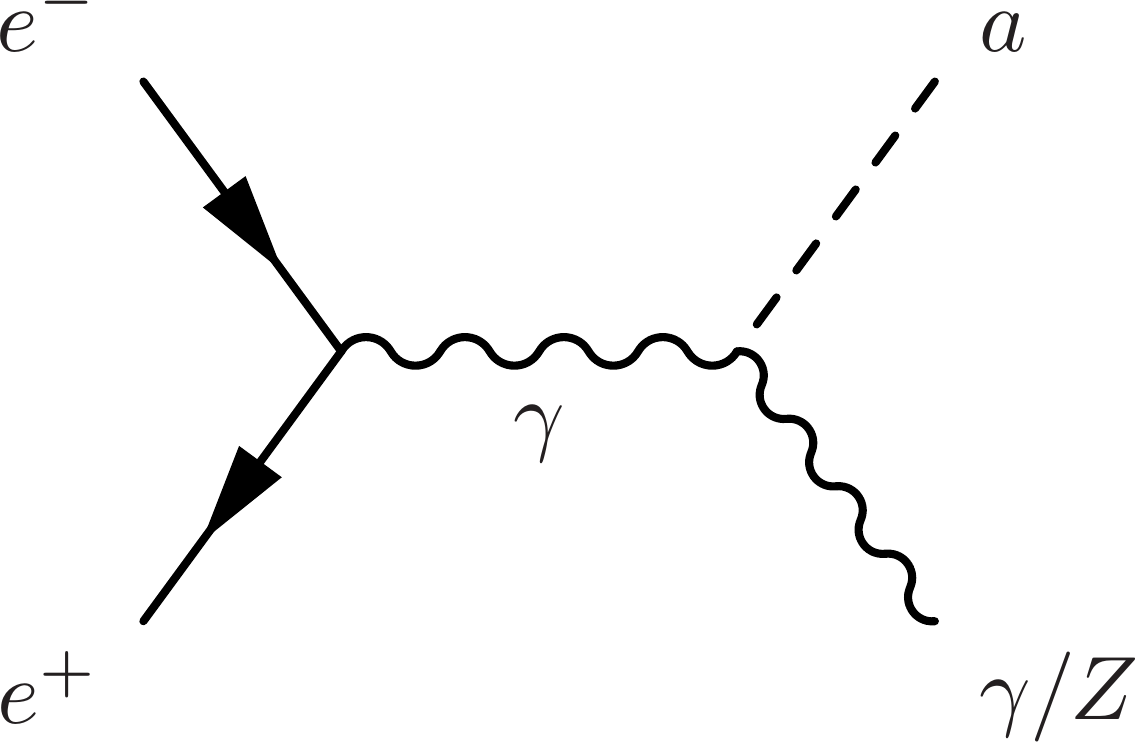} 
\caption{Tree-level Feynman diagram for the production of an axion in association with a photon or $Z$-boson.\label{FeynDiagram}}
\end{figure}

Explicit models relate these parameters to each other in model-specific ways and reduce the number of free parameters. One hereby generally expects the couplings to gauge bosons to be loop suppressed and of the same order, such that the gluon couplings dominate since $g_s\gg e$. However this does \emph{not} imply that a hadron collider is always the most sensitive machine: For \mbox{$m_a\lesssim100$\,GeV}, the QCD backgrounds at e.g.~the LHC are often simply too large, and the discovery mode could very well be through the electroweak couplings at the FCCee. Moreover, there exist models for which $c_{GG}\ll c_{WW}, c_{\gamma\gamma}, c_{\gamma Z}, c_{ZZ}$ \cite{Farina:2016tgd}, and for which a high energy lepton collider is the only viable probe. 
Specifically at the FCCee, ALPs can be produced either in exotic $Z$ decays (left panel of Fig.~\ref{FeynDiagram}) or in association with a photon or a $Z$-boson via an intermediate photon (right panel of Fig.~\ref{FeynDiagram}). The FCCee is expected to produce an unprecedented number of $10^{12}$ $Z$-bosons during its run at the $Z$-pole, $\sqrt{s} = m_Z$, which will let us search for extraordinarily small branching fractions for $Z\to a\gamma$ decays.
Once produced, the presence of an ALP can lead to different signatures inside the detector. ALPs can either be long-lived and travel through the detector unscathed or they can decay further into leptons, quarks or gauge bosons. Depending on their lifetime, ALPs may decay promptly at the interaction point or after they have travelled a certain distance inside the detector leading to a plethora of different signatures. 

\begin{figure}[t]\centering 
\includegraphics[width =0.32\textwidth]{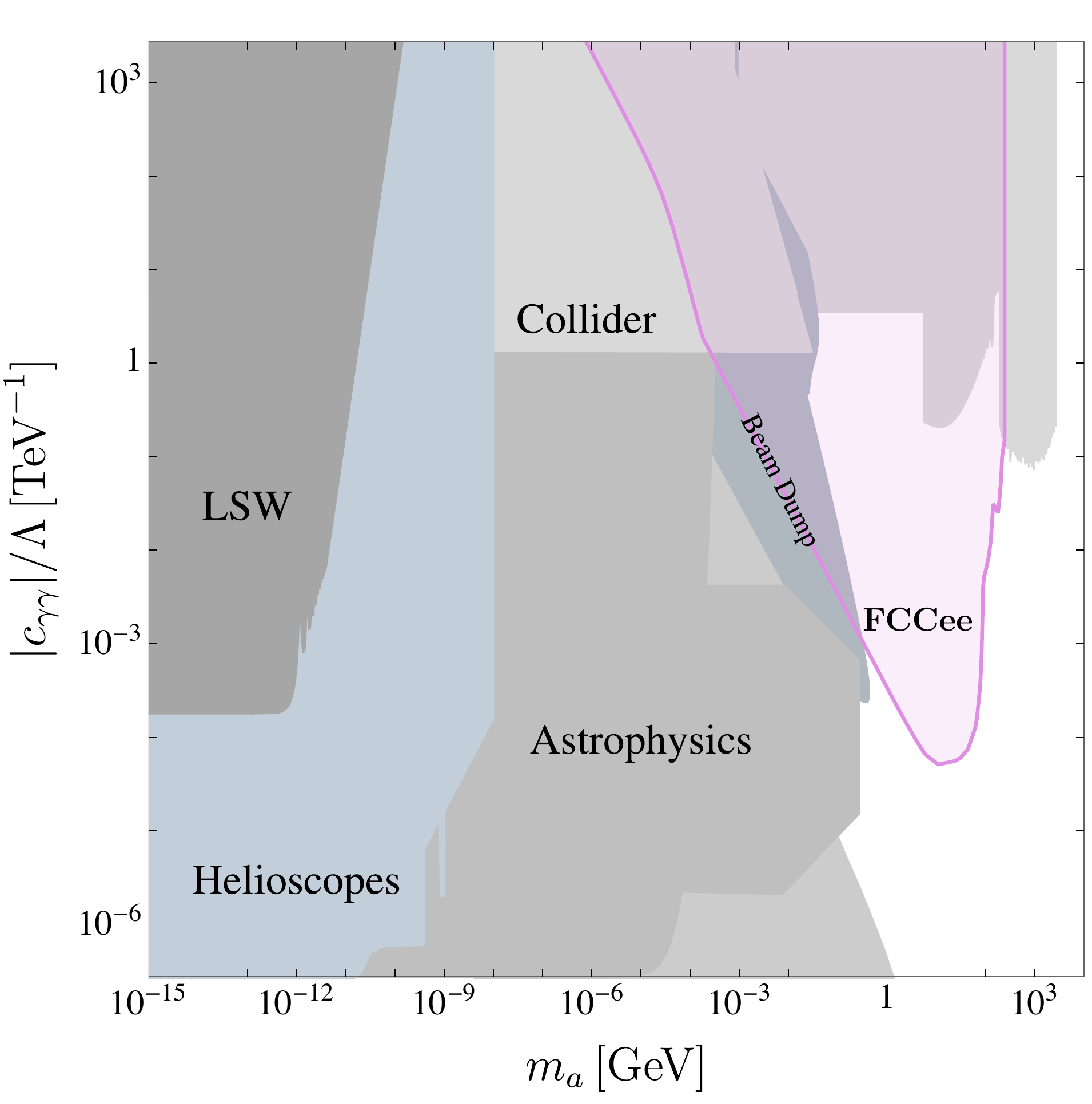}  \hspace{2cm}
\includegraphics[width =0.32\textwidth]{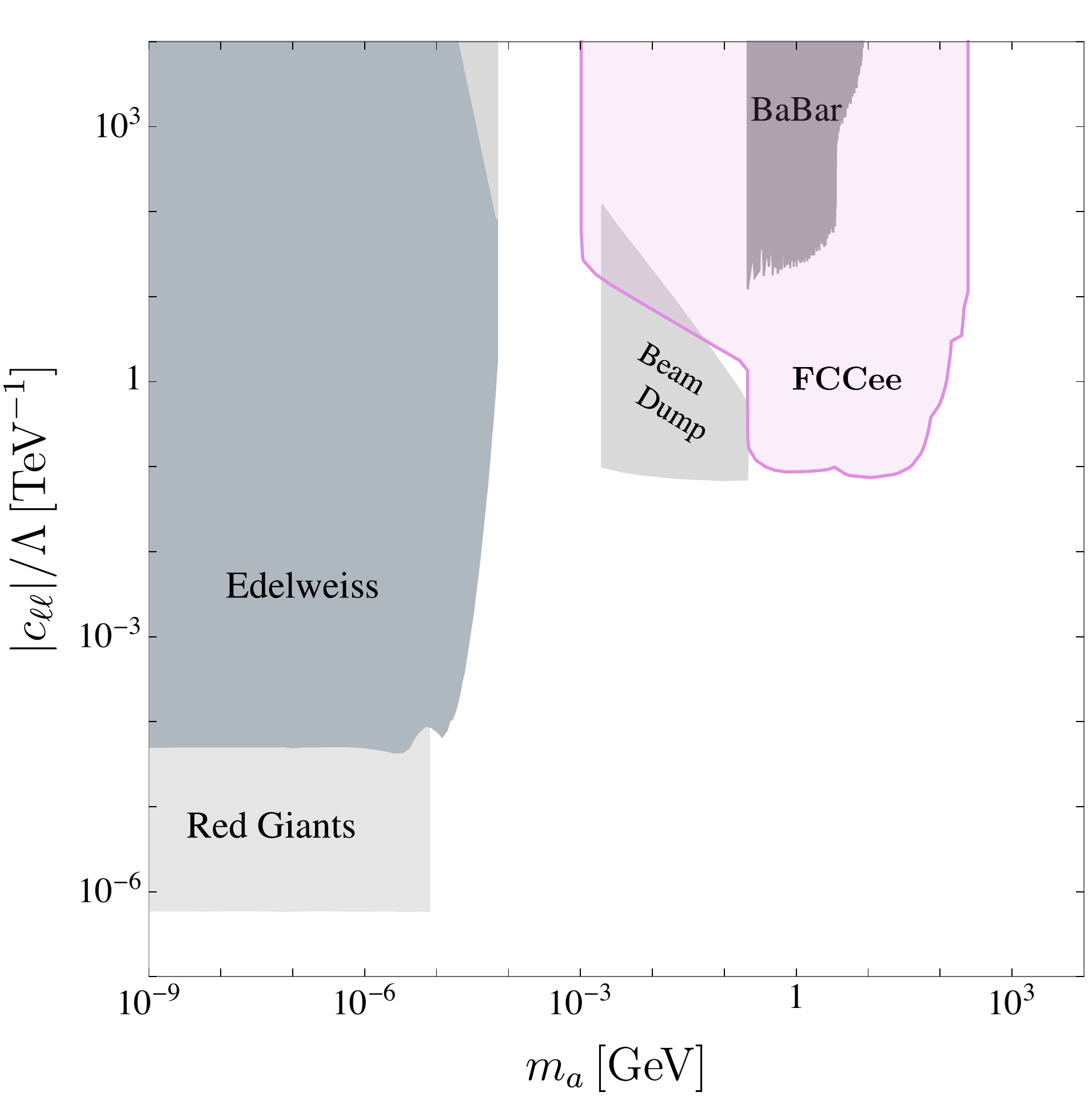}
\caption{Projected sensitivity of the FCCee (in purple) in the process $e^+ e^- \to \gamma a$ on the ALP-photon coupling (left) and the ALP-lepton coupling (right). Existing bounds on the parameter space are shown in grey. Reproduced from \cite{Bauer:2018uxu} with permission of the authors.\label{ALPreachPhotons}}
\end{figure}

The processes $e^+ e^- \to Z a \to Z \gamma \gamma$ and $e^+ e^- \to \gamma a \to 3 \gamma$ \cite{Liu:2017zdh,Bauer:2018uxu}, where the latter includes the production and decay of an on-shell $Z$-boson at the Z-pole, depend on the couplings $c_{\gamma\gamma}, c_{\gamma Z}, c_{ZZ}$, all of which can be related to each other in more concrete models. However, at the FCCee it is even possible to access $c_{\gamma\gamma}$ and $c_{\gamma Z}$ separately. The run at the Z-pole enhances the contribution of $c_{\gamma Z}$ to the process $e^+ e^- \to \gamma a$ with respect to $c_{\gamma\gamma}$ and thus $c_{\gamma Z}$ can be accessed at the Z-pole run while $c_{\gamma\gamma}$ can be measured at runs with a higher center-of-mass energy. Fig.~\ref{ALPreachPhotons} shows the parameter space that can be explored by the FCCee. Masses between hundreds of MeV and hundreds of GeV can be probed and the FCCee can push to very small values of $c_{\gamma\gamma}$.

The FCC also has great potential to probe the axion coupling to leptons, $c_{\ell\ell}$, which can be present in DFSZ type models. Interestingly, the dominant production mode at the FCCee is still in association with a photon or $Z$-boson where the ALP now couples to photons via a lepton loop. The ALP then decays to the heaviest lepton that is kinematically accessible. We show the expected sensitivity of the FCCee on the ALP mass and its coupling to leptons in the right panel of Fig.~\ref{ALPreachPhotons}.

In addition to direct measurements, the FCCee can probe electroweak precision observables and the electromagnetic coupling constant with unprecedented precision leading to further stringent constraints on $c_{\gamma \gamma}$ and $c_{\gamma Z}$.

\section{Heavy Neutral Leptons}

\begin{figure}[b]\centering \label{HNLreach}
\includegraphics[width =0.5\textwidth]{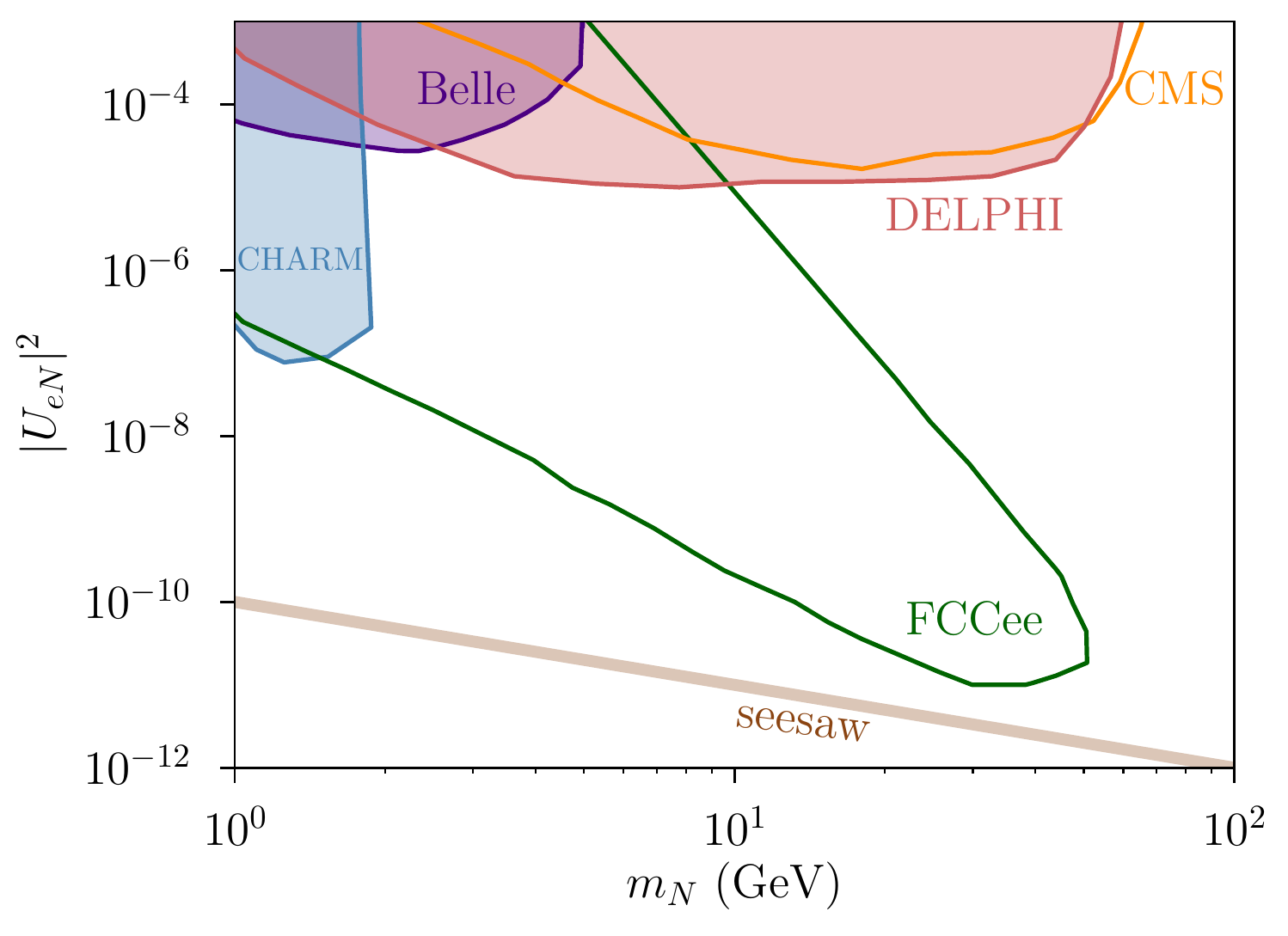}
\caption{Current limits from CHARM \cite{DORENBOSCH1986473}, Belle \cite{Liventsev:2013zz}, DELPHI \cite{delphi} and CMS \cite{Sirunyan:2018mtv} on a single HNL with mass $m_N$, compared with the sensitivity of the FCCee \cite{Blondel:2014bra}, assuming couplings to the first lepton generation only. The reach of the FCC on HNLs with couplings to the second and third lepton generation is similar. The brown band shows the prediction in a minimal seesaw model with a single HNL. It gives an indication of the expectation for the model in \eqref{eq:HNLlag} in the absence of a texture which further suppresses the active neutrino masses. }
\end{figure}

The (type I) seesaw mechanism \cite{Minkowski:1977sc,GellMann:1976pg,Freedman:1978gc,Yanagida:1979as,Glashow:1979nm,Mohapatra:1979ia} is perhaps the simplest and most natural mechanism explaining the smallness of neutrino masses in the SM, collectively denoted by $m_\nu$. It relies simply on a set of right-handed singlet fermions $N_{R,i}$, referred to as ``heavy neutral leptons'' (HNL) or ``sterile neutrinos'' with a Yukawa coupling, $y$, to the SM leptons $L$ as well as a Majorana mass $m$
\begin{equation}\label{eq:HNLlag}
\mathcal{L}\supset - y_{ij} \bar L_i \tilde H N_{R,j} - \frac{1}{2} m_{ij} \overline{N_{R,i}^c} N_{R,j} + \text{h.c.}\, , 
\end{equation}
where $\tilde H = i \sigma_2 H^*$. If $m_{ij}\gg m_\nu$, one can exchange the interactions in \eqref{eq:HNLlag} for Weinberg's effective interaction $y_{ik}m^{-1}_{k\ell}y_{\ell j} (\bar L_i \tilde H)(\tilde H L_j)^\dagger$. After substituting in the vacuum expectation value for the Higgs field $H$, this operator corresponds to a mass term for the SM neutrinos. Intriguingly, one obtains roughly the right mass range by setting $y\sim 1$ and $m\sim10^{16}\,$GeV, the scale at which grand unification ought to take place. 

On the other hand, $m$ could very well be much lower with $y\ll1$, such that the HNLs could a priori be in reach of our current or future accelerator facilities. In this context, it is useful to think of the active neutrino in the SM as a mixture of the light component and the sterile heavy neutrino
\begin{equation}\label{eq:massMixing}
\nu_{L,i} \approx \nu_i + U_{ij} N_{R,j}^c
\end{equation}
with an active-sterile mixing angle $U_{ij}^2 \sim m_\nu/m$. But as for the axion-like particles, there is a catch: For a single flavor, only the parameter space below the brown band in Fig.~\ref{HNLreach} would yield an acceptable mass for the active neutrino. Such small values of the mixing angle are likely not accessible with any near future facilities. 
This argument can be circumvented by noting that, to explain the masses of all three SM neutrinos, more than one HNL is needed. This allows for specific textures of $m_{ij}$ and/or $y_{ij}$ which suppress the contributions to the active neutrino masses relative to a ``generic'' $m_{ij}$ and $y_{ij}$. As a result, larger values of the mixing angle can be achieved without being in conflict with the stringent bound on the active neutrino masses. 

Alternatively, in a slightly less minimal setup, the HNLs could be Dirac states, where we add a lepton number preserving mass term $M_{ij} \overline N_{R,i} N_{R,j}$ to \eqref{eq:HNLlag}. If moreover $M \gg m$, the mass of the HNL mass eigenstate is given by $m_N\sim M$, while the SM neutrino masses are given by $m_\nu \sim y^2 m^2 /M$. By adding the extra knob of the lepton number-preserving mass $M$, we effectively decouple the mixing angles $U_{ij}$ from the SM neutrino masses and thus gain a large, more accessible parameter space, regardless of the need for specific textures \cite{Kersten:2007vk}. The price to pay is of course that this new model necessarily has more parameters than the minimal seesaw in \eqref{eq:HNLlag}, and a very large range of values for $U_{ij}$ are plausible. In some of the parameter space the model may moreover be responsible for the baryon and lepton asymmetry in the Universe \cite{Asaka:2005pn}.

During the Z-pole run of the FCCee, roughly $2 \times 10^{11}$ $Z$ bosons will decay into two left-handed SM neutrinos. If the SM neutrinos mix with the HNL, a large number of them may therefore be produced, even for exceptionally small mixing angles. \cite{Blondel:2014bra,Caputo:2016ojx,Liao:2017jiz,Antusch:2017pkq}. A further production mode is via  the higgsstrahlung process $Z \to Zh$ with the Higgs decaying into two HNLs \cite{Barducci:2020icf}.
At higher collider energies, the HNL production process will be dominated by the exchange of a $W$-boson. The HNL can decay via charged or neutral current processes, again through mixing with the SM neutrinos. In the former case, the flavor of the associated lepton will  contain information about the texture of $U_{ij}$.
The lifetime of the HNL varies strongly with $m_N$ and thus the decay length of a HNL can be substantial leading to a displaced vertex signature and reducing any background from $Z \to W^+W^-$ decays significantly. While light HNL are too long-lived to decay within the detector, the FCCee can set strong constraints on a HNL with masses larger than a few GeV and below the $Z$-mass, as shown in Fig.~\ref{HNLreach}. To leading order, the reach of the FCCee is independent of the flavor texture of the mixing vector $U_{iN}$ in \eqref{eq:massMixing}, though most existing constraints fairly sensitively depend on this. (See for a recent compilation of the bounds for different textures \cite{Beacham:2019nyx}.) However should an HNL be discovered, the FCCee will be able to pin down the different components of $U_{iN}$ by measuring the ratios of the $e$, $\mu$ and $\tau$ final states.

\section{Other opportunities}
To conclude, we briefly comment on a number of other opportunities at the FCCee. First and foremost, if the dark matter's origin is thermal and if its mass is below the mass of the $Z$ boson, we require a new mediator to set the correct relic density. Two well motivated examples are a scalar and a vector mediator which respectively mix with the SM Higgs and photon. A future $Z$-factory would be uniquely positioned to probe models of this kind  \cite{Liu:2017zdh}, since the missing energy signal at the LHC would typically be too low to be discoverable. 

Given the complexity of the SM model, it is moreover plausible that the dark sector itself also contains non-trivial, strong dynamics \cite{Strassler:2006im}. The GeV-scale is hereby especially preferred in the context of asymmetric dark matter \cite{Kaplan:2009ag}. Such a dark sector will generically be accessible through the $Z$-portal, providing another unique opportunity for a $Z$-factory such as the FCCee \cite{Cheng:2019yai}.

In recent years, a new solution to the hierarchy problem has emerged in which the expansion of the Universe during inflation is able to dynamically fine-tune the electroweak scale to a value much below the Planck scale \cite{Graham:2015cka}. Models of this kind predict a new, light scalar which mixes with the Higgs, the so-called \emph{relaxion}. A terra-$Z$ machine like the FCCee would be able to discover or exclude the relaxion for masses above $\sim 5$\,GeV \cite{Fuchs:2020cmm}.
 
Finally, there are still residual blind spots in more conventional frameworks such as supersymmetry: If the lightest neutralino is mostly bino-like and has a mass less $m_Z/2$, the exotic $Z$-decay $Z\to \tilde\chi_1^0\tilde\chi_1^0$ remains the best probe for this scenario. To prevent the neutralino from overclosing the Universe, it must decay through an R-parity violating coupling. The FCCee would be in an excellent position to find this type of decaying neutralino \cite{Wang:2019orr}.

\section*{Acknowledgments}
We are grateful to Prateek Agrawal, Michael Baker, Martin Bauer, Valerie Dompke, Jonathan Kozaczuk, Gaia Lianfranchi, Tongyan Lin, Matthew McCullough, Toby Opferkuch, Diego Redigolo, Ennio Salvioni and Kai Schmitz for useful discussions and suggestions when preparing this manuscript.

 \bibliographystyle{ejp} 
 \bibliography{FCCee}

\begin{thebibliography}{43}

\bibitem{Abel:2020gbr}
C.~Abel et~al. (nEDM), Phys. Rev. Lett. \textbf{124}, 081803 (2020),
  \texttt{2001.11966}

\bibitem{Peccei:1977hh}
R.~Peccei, H.R. Quinn, Phys. Rev. Lett. \textbf{38}, 1440 (1977)

\bibitem{Peccei:1977ur}
R.~Peccei, H.R. Quinn, Phys. Rev. D \textbf{16}, 1791 (1977)

\bibitem{Raffelt:1985nk}
G.G. Raffelt, Phys. Rev. D \textbf{33}, 897 (1986)

\bibitem{Raffelt:1987yu}
G.G. Raffelt, D.S. Dearborn, Phys. Rev. D \textbf{36}, 2211 (1987)

\bibitem{Lai:2002kf}
A.~Lai et~al. (NA48), Phys. Lett. B \textbf{536}, 229 (2002),
  \texttt{hep-ex/0205010}

\bibitem{Artamonov:2005ru}
A.~Artamonov et~al. (E949), Phys. Lett. B \textbf{623}, 192 (2005),
  \texttt{hep-ex/0505069}

\bibitem{Abouzaid:2008xm}
E.~Abouzaid et~al. (KTeV), Phys. Rev. D \textbf{77}, 112004 (2008),
  \texttt{0805.0031}

\bibitem{Ceccucci:2014oza}
C.~Lazzeroni et~al. (NA62), Phys. Lett. B \textbf{732}, 65 (2014),
  \texttt{1402.4334}

\bibitem{Rubakov:1997vp}
V.~Rubakov, JETP Lett. \textbf{65}, 621 (1997), \texttt{hep-ph/9703409}

\bibitem{DINE1981199}
M.~Dine, W.~Fischler, M.~Srednicki, Physics Letters B \textbf{104}, 199  (1981)

\bibitem{Zhitnitsky:1980tq}
A.~Zhitnitsky, Sov. J. Nucl. Phys. \textbf{31}, 260 (1980)

\bibitem{PhysRevLett.43.103}
J.E. Kim, Phys. Rev. Lett. \textbf{43}, 103 (1979)

\bibitem{SHIFMAN1980493}
M.~Shifman, A.~Vainshtein, V.~Zakharov, Nuclear Physics B \textbf{166}, 493
  (1980)

\bibitem{Bellazzini:2017neg}
B.~Bellazzini, A.~Mariotti, D.~Redigolo, F.~Sala, J.~Serra, Phys. Rev. Lett.
  \textbf{119}, 141804 (2017), \texttt{1702.02152}

\bibitem{Ferretti:2013kya}
G.~Ferretti, D.~Karateev, JHEP \textbf{03}, 077 (2014), \texttt{1312.5330}

\bibitem{Farina:2016tgd}
M.~Farina, D.~Pappadopulo, F.~Rompineve, A.~Tesi, JHEP \textbf{01}, 095 (2017),
  \texttt{1611.09855}

\bibitem{Liu:2017zdh}
J.~Liu, L.T. Wang, X.P. Wang, W.~Xue, Phys. Rev. D \textbf{97}, 095044 (2018),
  \texttt{1712.07237}

\bibitem{Bauer:2018uxu}
M.~Bauer, M.~Heiles, M.~Neubert, A.~Thamm, Eur. Phys. J. C \textbf{79}, 74
  (2019), \texttt{1808.10323}

\bibitem{DORENBOSCH1986473}
J.~Dorenbosch, J.~Allaby, U.~Amaldi, G.~Barbiellini, C.~Berger, F.~Bergsma,
  A.~Capone, W.~Flegel, L.~Lanceri, M.~Metcalf et~al., Physics Letters B
  \textbf{166}, 473  (1986)

\bibitem{Liventsev:2013zz}
D.~Liventsev et~al. (Belle), Phys. Rev. D \textbf{87}, 071102 (2013), [Erratum:
  Phys.Rev.D 95, 099903 (2017)], \texttt{1301.1105}

\bibitem{delphi}
D.~Collaboration, Zeitschrift f{\"u}r Physik C Particles and Fields
  \textbf{74}, 57 (1997)

\bibitem{Sirunyan:2018mtv}
A.M. Sirunyan et~al. (CMS), Phys. Rev. Lett. \textbf{120}, 221801 (2018),
  \texttt{1802.02965}

\bibitem{Blondel:2014bra}
A.~Blondel, E.~Graverini, N.~Serra, M.~Shaposhnikov (FCC-ee study Team), Nucl.
  Part. Phys. Proc. \textbf{273-275}, 1883 (2016), \texttt{1411.5230}

\bibitem{Minkowski:1977sc}
P.~Minkowski, Phys. Lett. B \textbf{67}, 421 (1977)

\bibitem{GellMann:1976pg}
M.~Gell-Mann, P.~Ramond, R.~Slansky, Rev. Mod. Phys. \textbf{50}, 721 (1978)

\bibitem{Freedman:1978gc}
D.~Freedman, P.~Van~Nieuwenhuizen, Sci. Am. \textbf{238}, 126 (1978)

\bibitem{Yanagida:1979as}
T.~Yanagida, Conf. Proc. C \textbf{7902131}, 95 (1979)

\bibitem{Glashow:1979nm}
S.~Glashow, NATO Sci. Ser. B \textbf{61}, 687 (1980)

\bibitem{Mohapatra:1979ia}
R.N. Mohapatra, G.~Senjanovic, Phys. Rev. Lett. \textbf{44}, 912 (1980)

\bibitem{Kersten:2007vk}
J.~Kersten, A.Y. Smirnov, Phys. Rev. D \textbf{76}, 073005 (2007),
  \texttt{0705.3221}

\bibitem{Asaka:2005pn}
T.~Asaka, M.~Shaposhnikov, Phys. Lett. B \textbf{620}, 17 (2005),
  \texttt{hep-ph/0505013}

\bibitem{Caputo:2016ojx}
A.~Caputo, P.~Hernandez, M.~Kekic, J.~L\'opez-Pav\'on, J.~Salvado, Eur. Phys.
  J. C \textbf{77}, 258 (2017), \texttt{1611.05000}

\bibitem{Liao:2017jiz}
W.~Liao, X.H. Wu, Phys. Rev. D \textbf{97}, 055005 (2018), \texttt{1710.09266}

\bibitem{Antusch:2017pkq}
S.~Antusch, E.~Cazzato, M.~Drewes, O.~Fischer, B.~Garbrecht, D.~Gueter,
  J.~Klaric, JHEP \textbf{09}, 124 (2018), \texttt{1710.03744}

\bibitem{Barducci:2020icf}
D.~Barducci, E.~Bertuzzo, A.~Caputo, P.~Hernandez, B.~Mele, JHEP \textbf{03},
  117 (2021), \texttt{2011.04725}

\bibitem{Beacham:2019nyx}
J.~Beacham et~al., J. Phys. G \textbf{47}, 010501 (2020), \texttt{1901.09966}

\bibitem{Strassler:2006im}
M.J. Strassler, K.M. Zurek, Phys. Lett. B \textbf{651}, 374 (2007),
  \texttt{hep-ph/0604261}

\bibitem{Kaplan:2009ag}
D.E. Kaplan, M.A. Luty, K.M. Zurek, Phys. Rev. D \textbf{79}, 115016 (2009),
  \texttt{0901.4117}

\bibitem{Cheng:2019yai}
H.C. Cheng, L.~Li, E.~Salvioni, C.B. Verhaaren, JHEP \textbf{11}, 031 (2019),
  \texttt{1906.02198}

\bibitem{Graham:2015cka}
P.W. Graham, D.E. Kaplan, S.~Rajendran, Phys. Rev. Lett. \textbf{115}, 221801
  (2015), \texttt{1504.07551}

\bibitem{Fuchs:2020cmm}
E.~Fuchs, O.~Matsedonskyi, I.~Savoray, M.~Schlaffer (2020), \texttt{2008.12773}

\bibitem{Wang:2019orr}
Z.S. Wang, K.~Wang, Phys. Rev. D \textbf{101}, 115018 (2020),
  \texttt{1904.10661}

\end{thebibliography}
%

%
%
%

%

\end{document}